\documentclass[pdflatex,sn-mathphys-num]{sn-jnl}
\usepackage[T1]{fontenc}
\usepackage{lmodern,amsmath,amssymb,booktabs,tabularx,graphicx,longtable,xcolor}
\hypersetup{hidelinks,bookmarksdepth=2}
\usepackage{microtype,enumitem}\setlist{nosep}\emergencystretch=1.5em
\newcommand{\code}[1]{\texttt{\small #1}}
\newcolumntype{Y}{>{\raggedright\arraybackslash}X}
\title{Operational limits of reference renewal for retained quantum-service results}
\author*[1]{\fnm{Roberto} \sur{\href{https://orcid.org/0009-0003-5312-2634}{Fernández-Barrios}}}\email{roberto.fernandez.b@deusto.es}

\author[1]{\fnm{Iker} \sur{\href{https://orcid.org/0000-0002-3068-6248}{Pastor-López}}}\email{iker.pastor@deusto.es}

\author[1]{\fnm{Amaia} \sur{\href{https://orcid.org/0009-0003-9080-6242}{Pikatza-Huerga}}}\email{a.pikatza@deusto.es}

\author[1]{\fnm{Pablo} \sur{\href{https://orcid.org/0000-0003-3594-9534}{García Bringas}}}\email{pablo.garcia.bringas@deusto.es}

\affil[1]{\orgdiv{Faculty of Engineering}, \orgname{University of Deusto}, \orgaddress{\street{Avda. de las Universidades 24}, \city{Bilbao}, \postcode{48007}, \country{Spain}}}
{}
\begin{document}
\abstract{
We implement reference renewal to reconsider retained physical quantum-service results under a prespecified protocol. Twelve transitions retained six results each; ten completed renewal, eight met the prespecified endpoint, and five succeeded using retained results alone. The exact one-sided $p=0.0546875$ does not establish the recovery hypothesis. Incomplete references, binding failures and sequential waits expose operational limits. Offline reconstruction recovers decisions from preserved counts. Retrospective binding-plus-oracle checking accepts all 60 usable-transition results without renewal. With the exact oracle also supplying calibration scores, the study does not demonstrate incremental numerical-correctness utility of renewal.
}
\maketitle

\paragraph{Acknowledgments and funding.} This work is part of grant PID2024-155693NB-C43, ATHENA-AEGIS (Advanced Secure Technologies for Hybrid Quantum-Classical Environments and Applications), funded by MICIU/AEI/10.13039/501100011033 and by ERDF/EU. MICIU and AEI denote the Ministerio de Ciencia, Innovación y Universidades and the Agencia Estatal de Investigación, respectively; ERDF/EU denotes the European Union through the European Regional Development Fund.
\clearpage

\section{Introduction}
We implement and evaluate a procedure that reconsiders retained physical quantum-service results after renewal of their supporting reference. Its operational sequence binds each result to the observed execution context, acquires a complete replacement reference, re-evaluates the same result, and records its source and service deadline. Quantum software engineering treats operation, quality assurance and demand/capacity management as connected concerns \cite{murillo2025roadmap}; this study examines their interaction in one executed procedure, with reproducible decisions and an explicit direct-checking comparison.

The physical record contains twelve transitions with six bound results retained in each. Ten completed reference renewal; eight of those ten met the prespecified endpoint, and five did so using retained results alone. The two incomplete transitions, binding disagreements and long sequential waits show where the implemented reference requirement affects service completion. These observations are distinct from facts fixed by the rules, such as requiring 39 calibration jobs for a complete 39-unit reference. The offline artifact reconstructs the outcomes from preserved counts and separately evaluates whether the same physical results satisfy binding and absolute-error checks without renewal.

The physical result, its appraisal reference and the service decision are different objects. In this sequence the probe is acquired under the new observed context while the available reference belongs to an earlier context. Recalibration changes that reference; it does not change the retained measurement. Missing execution evidence cannot be repaired by a fresh reference, and a classical fallback changes the source of the answer. Whether renewal is useful beyond direct checking, and whether a different scheduler would meet the deadline, require evidence beyond the rule-defined restoration of eligibility.

Fernández-Barrios, Pastor-López, Pikatza-Huerga and García Bringas, \emph{Observational Indistinguishability and Integrity Blind Regions in Hybrid Quantum-Classical Workflows} \cite{fernandez2026observational}, supply the claim-relative evidence/reference perspective. That predecessor develops an integrity-view lattice, structural blind regions, reference sufficiency, an offline policy analysis and a simulator-only quantum branch. It explicitly excludes physical quantum processing unit (QPU) and deployed-service assurance. We inherit the discipline of distinguishing claims from observations, the kernel-workload lineage, calibrated decisions and evidence contracts. We do not claim its lattice, the conformal ranking rule, abstention or authenticated provenance as new results.

This work is part of ATHENA-AEGIS (Advanced Secure Technologies for Hybrid Quantum-Classical Environments and Applications). Within that project, the evaluated component is the retained-result assurance procedure and its local service runner. The experiment does not validate the complete project architecture or an industrial application.

The implemented sequence uses established monitoring, calibration and provenance ideas. The empirical contribution is its observed completion and failure behavior under a locally prespecified protocol, rather than a claim of a new assurance pattern. Recorded waits, incomplete references and late service are execution outcomes; the reference size and the eligibility predicate are design choices. Comparative deployment latency, monetary or energy savings, and an external necessity for renewal were not evaluated.

For this two-qubit workload, the exact classical answer is already available before acquisition and supplies every nonconformity score. A simpler alternative can therefore preserve the very same physical result, check its binding and test absolute error directly, without renewing a reference. Our retrospective reconstruction accepts all 60 usable-transition probes under that alternative. The reference-renewal procedure restores eligibility under its own contract, but this observation does not establish an external need for that contract or additional numerical utility. Because the same exact oracle already determines the absolute-correctness target and every score, this design cannot demonstrate a numerical-correctness advantage of renewal over direct checking; collecting more windows of this kind would not remove that limitation.

The contribution is a bounded implementation study with count-based reconstruction, operational accounting and a retrospective direct comparator. It specifies how binding, reference support, numerical correctness, source and deadlines interact, and reconstructs the observed decisions from preserved measurement counts. The prespecified endpoint and the retrospective direct comparator answer different questions: the former measures completion under the renewal contract; the latter tests whether the same correct physical result could be admitted without renewal. Neither analysis establishes comparative superiority or a reason to prefer quantum acquisition.

\section{Related work and comparative scope}
Runtime verification checks observations of executions against specified properties \cite{leucker2009brief}; a monitor's verdict and the service response to it are separate design choices. Slagel et al. formalize Simplex runtime-assurance switching through proof obligations in PVS and differential dynamic logic \cite{slagel}; our software has no comparable machine-checked safety theorem. SelectiveNet learns prediction and rejection jointly \cite{selective}, whereas our admission predicate concerns evidence support rather than a learned risk--coverage tradeoff. Conformal methods can address departures from exchangeability under additional assumptions \cite{barber}; equality of our context digest does not supply such a correction.

Quantum assertions already inspect properties of program states. Li et al.'s Proq represents predicates as projections and checks them through projective measurements \cite{li2020projection}. RuntimeSpeQ specifies states and execution steps and translates specifications into quantum assertions \cite{runtimespeq}. Its experiments use manual translation followed by automated checking with Qiskit Aer; the evaluation concerns fault detection, equivalence checking and specification reuse. Rovara et al. translate assertions into measurement-bearing program slices, apply optimizations intended to reduce execution overhead and statistically evaluate the results under noise \cite{rovara}. The reported experiments include IBM hardware and specify their shot budgets. Thus hardware evaluation and measurement-aware monitoring are established antecedents. Our checked object is instead the eligibility of a retained result under a renewed reference, followed by a source-and-deadline service decision. Table~\ref{tab:comparison} summarizes the closest procedural distinctions.

QCIVET already combines calibrated observable contracts, hash-linked traces, conditional verification properties and hardware evaluation \cite{yeniaras2026qcivet}. QML-PipeGuard studies channel substitution, blind sets and benign drift with finite-shot analysis and hardware experiments \cite{yeniaras2026pipeguard}. We compare its public extended v2, whose record reports a shorter paper accepted at NordSec 2026; acceptance of that paper is not evidence of peer review of every extended result. The cloud-auditing example in QCIVET also includes backend/calibration checks and a stale-snapshot tampering case. We do not claim contracts, execution-context binding, authenticated provenance or quantum monitoring as new. Our evaluated endpoint is a six-probe recovery window with retained-versus-classical source restrictions and a deadline, rather than attack detection or channel discrimination.

Observability and provenance also have established workflow antecedents. Beisel et al. describe process views spanning classical services, infrastructure and quantum-device characteristics \cite{beisel}. Weder et al. combine workflow rewriting for hybrid runtimes with provenance collection and views that retain the connection to the modeled workflow \cite{weder}. Peltonen et al. distinguish evidence origins and unavailable fields in a cross-provider API, evaluated with a fixture-backed adapter prototype \cite{peltonen2026provenance}. Our historical-context predicate uses selected observations for a campaign-specific admission decision; it neither replaces those broader observability tools nor supplies stronger independent processor attestation.

Physical drift detection uses time-resolved circuit observations and spectral analysis in Proctor et al. \cite{proctor}. We test customer-side metadata consistency and reference renewal without identifying a physical drift mechanism. Ravi et al. analyze cloud workloads, queueing and calibration crossovers \cite{ravi}; these operational phenomena are not discoveries of this study. QOS evaluates resource management on real devices \cite{qos}, while Qurator schedules hybrid tasks using fidelity and queue estimates with a trace-driven evaluation \cite{pehlivanoglu2026qurator}. Our fixed sequential policy is not a scheduling optimizer.

The recovery layer of Siddi Moreau et al. is a closer comparison \cite{workflow}. It combines retries, speculative replicas and backend failover with quorum aggregation, warm starts and run-level resumption from checkpoints. Its provenance records distinguish attempts and recovery actions. The reported ordinary operation absorbs losses through quorum/warm-start behavior without redispatch; the resilience study exercises re-execution under injected failures, and a separate device breakdown includes real QPUs. These mechanisms continue a solver workflow using completed samples, replacement attempts or saved solver state. Here, the held physical evidence is already bound to the new context: only its supporting reference is replaced, and that same evidence is reconsidered without another service-probe execution. The consulted workflow paper does not evaluate this specific reference-renewal endpoint. This is a distinction between the procedures evaluated, not a claim that its architecture could not incorporate ours, or that our policy is preferable.

\begin{table}[htbp]\centering\footnotesize
\begin{tabularx}{\linewidth}{p{.16\linewidth}YYY}\toprule
Work & Checked object / evidence & Subsequent action & Evaluated outcome\\\midrule
RuntimeSpeQ \cite{runtimespeq} & State and step specifications; assertion observations & Check translated monitors & Fault detection, specification-relative program equivalence and reuse in Aer\\
Rovara et al. \cite{rovara} & Program assertions; measured slices and noise estimates & Statistical assertion verdict & Assertion detection on hardware at reported shot budgets\\
Siddi Moreau et al. \cite{workflow} & Subproblem/run status; samples, attempt records and solver state & Quorum/warm-start continuation; retry, replica, failover or checkpoint resumption & HPC orchestration and injected-failure resilience; device latency profiles\\
This study & Bound held result; complete context-matched reference and oracle score & Re-evaluate retained evidence; serve retained result or permitted fallback & Timely six-probe recovery: 8/10, H1-B not established\\\bottomrule
\end{tabularx}
\caption{Objects, actions and reported evaluations in the consulted versions. RuntimeSpeQ is read through its author manuscript and Rovara et al. through arXiv:2505.03885v1 with published metadata. The specific retained-result reference-renewal endpoint is not evaluated in those works; this does not imply lack of architectural support.}\label{tab:comparison}
\end{table}

These comparisons are not head-to-head performance findings, a proof of priority or a systematic review. Online Resource 1, Section S1, distinguishes the predecessor preprint \cite{fernandez2026observational} and the authors' other work. Neither the predecessor's structural integrity results nor the authors' certificate and acquisition-cost theorems are claimed again. The present contribution is the implemented evidence-to-service sequence and a bounded evaluation of its recovery consequences.

\section{System, context and decision model}
\subsection{Observable context and trust boundary}
The ATHENA-AEGIS implementation evaluated here is a local runner around a cloud quantum backend. A unit produces a result $r$, execution evidence $e$ and an observable workload projection $c=\Phi(e)$. The frozen projection includes the backend, physical qubits and used operations, relevant calibration values and target duration/error values, layout, transpilation settings, shots, runtime options, software versions, workload/data identity and circuit semantic hash. Per-parameter dates are excluded from $\Phi$ to avoid equating every backend-wide timestamp change with a workload-relevant transition; full snapshot identity and the window calibration timestamp remain separately recorded. Reference support means exact equality of the declared stratum, including the workload-context digest, between the calibration units and the held result.

A retrospective description compares each usable window with the stale reference actually selected by the procedure, not necessarily the immediately preceding candidate (Supplement S8.6, Table~S9). The ten comparisons contain 2--24 changed scalar JSON leaves, all in calibration-property views. For C08/W06 versus C07, only the provider's measurement \texttt{threshold} values for qubits 0 and 1 differ; these are not the scientific acceptance or correctness thresholds. Repeated representations of a parameter are separate JSON fields, not independent physical changes. The contract treats the observed context change as invalidating prior reference support; neither a digest change nor its field count establishes material error-distribution drift, physical irrelevance or a causal effect on the results. Unchanged observed fields cannot establish equality of the complete physical state.

Let $B(e,c)$ denote the conjunction of four binding checks: J1 requires provider DONE and monotonic provider timestamps; J2 requires the pre-submit and post-result local snapshots to carry the window digest; J3 requires the historical workload projection at the provider running timestamp to equal that digest; J4 requires result integrity. A failed or unknown required check makes the unit audit-only. In particular, agreement of local snapshots alone is insufficient.

The frozen collector captures backend status, local properties and historical properties indexed by provider execution time. These are the recorded observations used by the predicate, not an assumed post-submit calibration lock. We make no claim about capabilities of a later provider interface. Historical metadata can be incomplete or inaccurate; the service does not independently observe the physical processor's complete state. The trusted boundary comprises the local collector, oracle, clock, integrity mechanisms and the provider fields used by binding. The campaign does not establish resilience to a malicious provider or a compromised local signer.

\subsection{Workload and correctness}
The physical evaluation uses \code{paper15-kernel-regression-v1} on \code{ibm\_basquecountry}, physical qubits $[0,1]$, layout $[0,1]$ and a shot setting of 128 for each circuit. The feature points are $(0.1,0.2)$, $(0.3,0.4)$ and $(0.5,0.6)$. Following the quantum-feature-space kernel construction of Havl{\'i}{\v{c}}ek et al. \cite{havlicek2019supervised}, a two-repetition ZZ feature map and compute--uncompute circuits estimate the six upper-triangular entries of a symmetric $3\times3$ fidelity kernel from all-zero measurement probabilities. A physical job/unit therefore contains multiple circuits; job counts are not circuit counts. The frozen implementation uses Qiskit's \texttt{zz\_feature\_map} and composes each data-encoding circuit with the inverse encoding for its paired point (Supplement S7.1). This attributes the construction, not a reproduction of the cited classifier or its performance claims. Despite the workload identifier, the scored service artifact is the kernel matrix, not a downstream regression prediction. The exact-statevector reference makes numerical correctness decidable for this small workload. The nonconformity score is relative Frobenius error of the measured kernel against that reference; the oracle also checks maximum absolute error, with frozen thresholds 0.15 and 0.20 respectively and a $10^{-12}$ numerical comparison tolerance. These are tolerance-based numerical criteria, not exact equality of a finite-shot kernel and the statevector kernel.

An oracle-correct answer and assurance-valid serving are different predicates. The former concerns numerical service quality. The latter means that the implemented appraisal has its required binding, reference and support premises. It is an operational eligibility flag, not a newly established statistical coverage theorem. Because the exact reference is available both for scoring and for fallback, the study does not solve the oracle problem for classically intractable quantum computation. A classical-only service is a serious alternative on this workload.

\section{Implemented recovery procedure}
\subsection{Reference renewal and retained-result reuse}
For a complete reference $\mathcal C_c$ of 39 context-bound calibration scores, the frozen rank rule uses the standard conformal nonconformity ranking construction \cite{shafer2008tutorial}:
\begin{equation}
 p_c(s)=\frac{1+\sum_{a\in\mathcal C_c}\mathbf{1}\{a\geq s\}}{40},\qquad \text{reject if }p_c(s)\leq0.05.
\end{equation}
The tie convention counts calibration scores at least as large as the test score. With 39 calibration scores, the rule rejects exactly when no more than one calibration score is at least the test score. This threshold is not an empirically validated false-rejection rate in this study. Every calibration score, as well as the held-probe score, is computed against the exact classical matrix. Reference renewal itself therefore uses the oracle; this dependence is not confined to evaluation or fallback. A setting without that reference would require a different score construction, which is not evaluated here. Its usual calibration interpretation requires conditional within-window exchangeability of calibration and eligible test scores. The exchangeability caveat is substantive \cite{barber}. Neither hashing the context nor a complete 39-unit reference proves that premise; the sequential acquisition order and shared device leave it an assumption.

\begin{figure}[htbp]\centering
\includegraphics[width=\linewidth]{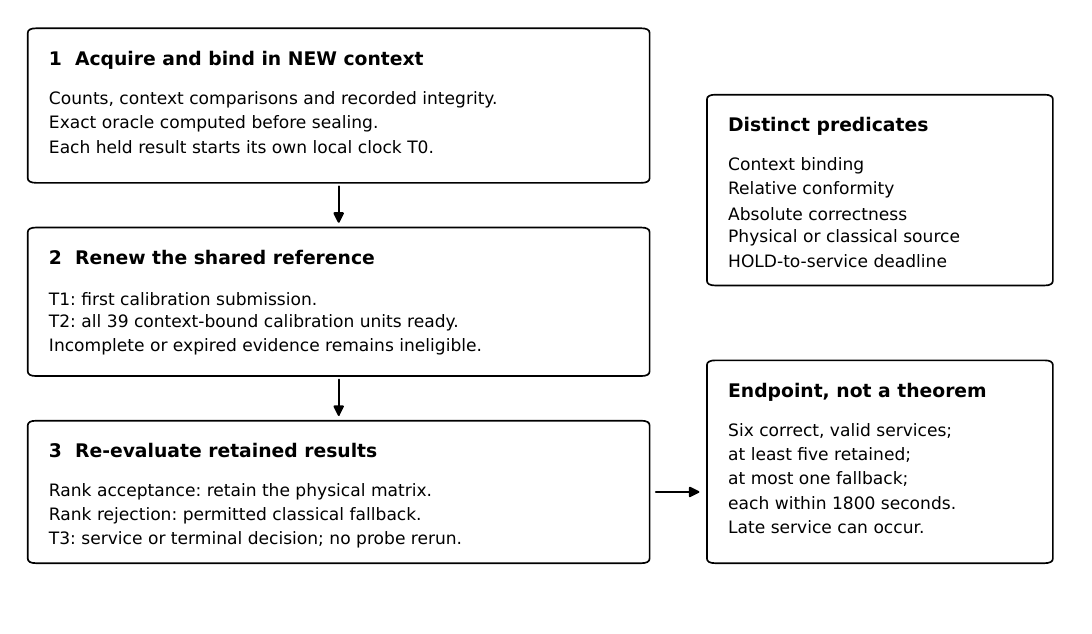}
\caption{Implemented recovery path and evaluation rule. The deadline is an endpoint criterion, not a hard service cutoff: late correct, valid service still fails the primary endpoint.}\label{fig:mechanism}
\end{figure}

P2 applies the most recent prior accepted, complete, bound reference to the new context without the explicit support guard. If none exists, it is calibration-invalid. P4 applies the support guard and withholds an unsupported result. P5 adds the executed recovery sequence in Fig.~\ref{fig:mechanism}. Each of six service probes is acquired once and held on sealing. Thirty-nine calibration jobs then produce the new reference. P5 re-evaluates the held results without rerunning them. An admitted score permits serving the retained result; a rejected score permits the declared classical fallback only after the physical result, binding, complete reference and re-evaluation exist. Missing physical evidence, failed binding, an unknown oracle state or an incomplete reference cannot be rescued by fallback.

The code admits fallback per probe; the window endpoint requires at most one fallback and at least five retained results. This restriction prevents an all-classical window from satisfying the primary claim. The contract permits at most one acquisition retry per window after an eligible terminal failure or submission exception before held evidence exists; rerunning after HOLD is forbidden. The sole recorded retry was a C01 calibration submission; both attempts failed before provider acceptance (Supplement S3.3). It was not a repeated physical execution. Retrying acquisition and re-evaluating existing evidence are different actions.

Crucially, the retained result is acquired in the \emph{new} window and bound to its context; it is the prior calibration reference that is stale. This procedure does not migrate an old-context physical result into a new context merely by changing its certificate. Context validity must hold for both the retained evidence and the replacement reference.

P2 and P4 are policy replays on the same physical probe evidence, while P5's recalibration is physically executed. They are mechanistic comparators, not separately randomized service deployments. Campaign B's original protocol specifies a one-sample P5 window test, not a confirmatory P5-versus-P2 treatment-effect test.

\subsection{Predicates and a direct comparator}
Binding asks whether the available execution evidence matches the declared context. Conformity asks where the score lies among 39 reference scores. Absolute correctness separately requires relative error at most 0.15 and maximum absolute error at most 0.20. Source records whether the output is retained physical evidence or classical fallback. The deadline evaluates elapsed time from HOLD. These predicates are not interchangeable.

In the executed P5 path, binding and reference support gate reconsideration, and the rank rule selects retention or fallback. Absolute correctness and deadline compliance are checked in the endpoint analysis; neither is a general runtime guarantee. For example, 39 calibration scores of 0.30 and a test score of 0.20 yield rank value 1 and acceptance, although the test score exceeds the absolute relative-error threshold of 0.15. The frozen rank implementation confirms this synthetic software case (Supplement S8); it is not a Campaign B error.

The retrospective direct comparator requires a physical result, the same available binding evidence and both absolute oracle thresholds, and returns that physical matrix only when all are satisfied. It neither renews a reference nor substitutes the classical matrix. Missing required evidence yields an explicit unknown unless another required predicate has a known failure, which takes precedence as rejection. Recorded absence of physical execution is such a failure; unknown execution status is not. The frozen runner computes the oracle and binding before sealing each unit. For all 72 held probes, sealing precedes HOLD, so the required numerical and recorded binding information was available by that event. This establishes retrospective decision feasibility, not measured comparator response time. Recorded integrity checks and provider metadata remain shared trust premises.

\subsection{Clock, scientific termination and endpoint}
For probe $i$ in window $w$, $T_{0,wi}$ is its sealed-result HOLD decision, $T_{1,w}$ the first calibration submission, $T_{2,w}$ completion of the context-bound reference, and $T_{3,wi}$ service or terminal rejection. The recovery clock is $L_{wi}=T_{3,wi}-T_{0,wi}$. It includes subsequent waiting and sequential work, including acquisition of later held probes. It excludes acquisition time before that probe's HOLD. Provider queue components are reported separately and never subtracted from $L$.

Define $R_{wi}$ as the conjunction of bound held physical evidence, served output, an oracle-correct output with a determinate verdict, valid assurance, $L_{wi}\leq1800$ s, no service-probe rerun, the recorded binding-invariant flag, and a known source in \{retained result, classical fallback\}. The window endpoint is
\begin{equation}
Y_w=\mathbf{1}\left\{\bigwedge_{i=1}^{6}R_{wi},\;N_{\rm retained}\geq5,\;N_{\rm fallback}\leq1,\;\text{six distinct probes}\right\}.
\end{equation}
In the frozen producer, \texttt{trusted\_invariants\_held} is assigned from the unit's \texttt{binding\_valid} flag; the runner verifies the local HMAC it has just created. The flag defaults to true even for some failed or unbound attempts, so by itself it does not establish physical execution or successful J1--J4 binding. It is a local self-check, not an independent safety attestation, and is an explicit trust premise of the direct comparator. Throughout, ``served'' denotes the runner's recorded service decision, not measured delivery to an external end user.

The primary safety gate rejects a success declaration for any served primary P5 record with incorrect/unknown oracle, invalid/unknown assurance, failed/unknown binding or invariant failure. This is an implemented evidence check over observed records; zero observed violations is not a population safety certificate.

Context expiry or 24 hours after window opening terminates the scientific window irrevocably. Later results are audit-only and cannot repair eligibility. Provider QUEUED is not itself execution failure; running jobs are not cancelled by the expiry rule. Incomplete scientific records remain incomplete even if a provider eventually supplies a result. These rules prevent waiting and reconciliation from silently changing the analysis set.

\section{Prespecified experimental design}
\subsection{Developmental antecedent and frozen Campaign B design}
Campaign A had its own locally fixed protocol and remains a closed study under that contract. It produced five usable transitions, below its required minimum of eight; H1-A was not established and its primary inference was not computed. In those five windows, neither P2 nor P5 served an oracle-incorrect output (0/30 each). Its 31 local timeouts, three incomplete epochs and three accepted non-usable transitions exposed evidence-acquisition problems. A's operational metadata informed B's admission gate; the absence of observed incorrect service motivated a recovery endpoint rather than a repeated claim of error prevention. These are documented design motivations, not a causal before/after evaluation. Table~\ref{tab:campaigns} summarizes the contracts; Supplement S2 retains A's result, late-result exclusions and retrospective field clarification.

\begin{table}[htbp]\centering\small
\begin{tabularx}{\linewidth}{p{.20\linewidth}YY}\toprule
 & Campaign A: antecedent & Campaign B: principal evaluation\\\midrule
Primary question & Paired P5--P2 unsafe-service difference & Six-probe timely valid recovery\\
Required evidence & At least eight usable transitions & Exactly ten usable transitions\\
Observed flow & 12 opened; one bootstrap; five usable; three accepted non-usable; three incomplete & 14 opened; two bootstrap attempts (one completed); ten usable; two incomplete transitions\\
Job waiting & Local 3,600 s timeout & Provider terminal states; context expiry or 24 h scientific limit\\
Context evidence & A's original context rules & Explicit J1--J4 execution binding\\
Result & Minimum unmet; H1-A not established & 8/10; $p=0.0546875$; H1-B not established\\\bottomrule
\end{tabularx}
\caption{Separate prespecified campaigns. A and B differ in endpoint, eligibility, job handling and time period. Samples are neither pooled nor used to estimate a causal protocol effect.}\label{tab:campaigns}
\end{table}

The protocol was fixed locally in a version-controlled contract with annotated Git tags and documented execution amendments (Supplement S3). No independent pre-acquisition timestamp anchor or external registration for that contract has been established from the available record. We therefore describe the study as conducted under a prespecified protocol; local commit timestamps alone do not independently attest precedence. Historical filenames and tags retain their original names. A B candidate requires an operational backend with status active, queue depth at most five, executable fixed support, a genuinely new workload-context digest and a strictly later provider calibration timestamp. Two qualifying snapshots must be at least 600 s apart; the gate must also hold immediately before opening. After opening, queue depth does not throttle scientific units. A quiet admission snapshot is not a reservation of future queue capacity.

Within the fixed stage order, the frozen design assigns block-wise randomized units and replay order using a committed seed and CPython 3.11 shuffle. Acquisition remains sequential: six held probes, 39 calibration jobs, zero-QPU re-evaluation and two secondary verification jobs. Bootstrap has 39 calibration and two verification jobs, with no primary probes. A usable transition requires all six primary probes to have exactly one bound sealed attempt, a complete bound reference, re-evaluation before expiry and integrity checks. Verification-probe absence alone does not void a usable window.

\subsection{Population, stopping and inference}
The sample target was ten usable transitions, with a maximum of 16 candidate epochs and other frozen limits: 810 budgeted jobs, 2,430 conservative QPU seconds, quota guards, fatal criteria, 90 calendar days or 30 days without a gate-open opportunity. Stopping is based on usable-window count and operational limits, not the accumulating success count. Three documented amendments, one authorized before acquisition and two during execution, address digest scope, the provider request boundary and launch-record contract provenance. They retain hypotheses, endpoints and thresholds. The closure dossier documents the amendment chain; this manuscript neither re-audits nor changes it.

The statistical model used by the prespecified binomial test treats ten windows as independent Bernoulli observations with common probability $\pi_B$ and tests $H_0:\pi_B\leq0.5$ against $H_{1\text{-B}}:\pi_B>0.5$ at one-sided $\alpha=0.05$. Rejection requires $X=\sum_wY_w\geq9$. Windows, not their six nested probes, are replicates. If fewer than ten usable windows exist, the rule is not tested. Secondary endpoints are descriptive and cannot be promoted when H1-B is not supported.

The stated target population is gate-admitted genuine transitions. In practice, inference uses only those also satisfying usability, which requires successful evidence completion before expiry. This selection is outcome-blind with respect to the final numerical verdict, but can still select favorable operational conditions. Consequently, the observed contrast characterizes the usable, gate-admitted subset; extrapolation to all gate admissions requires additional assumptions about incomplete windows. We preserve the denominator ten and disclose the failed bootstrap attempt and the two incomplete transitions without treating them as primary failures or omitting their accounting.

\subsection{Retrospective reconstruction and design sensitivity}
The local revision record places an explicit retrospective specification before the new calculations; this ordering has no independent timestamp attestation. They reconstruct errors, old/new ranks, joint decisions, reference ages and the direct comparator from the existing records. The ten-window confirmatory set, thresholds and endpoint remain unchanged; C05/C09 are a separate descriptive traversal. Synthetic semantic and analyzer regression tests are software checks, not physical observations.

With ten independent homogeneous windows and rejection at nine successes, power is 0.375810 at hypothetical $\pi_B=0.8$ and 0.736099 at $\pi_B=0.9$. The actual size at $\pi_B=0.5$ is 0.010742, below the nominal 0.05 because the rejection region is discrete. These calculations describe design sensitivity; they neither explain the observed outcome as a power failure nor reveal a hidden advantage. The historical design report documents the author's ten-window target, an operational epoch/calendar envelope, and the choice of the 0.5 null to accompany the nine-success rule. It does not establish a utility threshold or a mathematically minimal sample size (Supplement S8).

\section{Results}
\subsection{Flow and confirmatory outcome}
Fourteen candidates opened, including two bootstrap attempts: C01 halted before provider acceptance and C02 completed bootstrap. The operational path is twelve transitions with retained results, ten complete renewals and usable windows, and eight successes within the deadline. Each transition acquired six bound physical results; C05 and C09 expired with their six probes held and only 30/39 and 22/39 bound calibrations, respectively. Eight of the ten usable windows met the primary endpoint; five met the stricter retained-only endpoint (all six timely, correct and valid, with no fallback). The latter is descriptive, not another confirmatory contrast. No accepted non-usable window occurred, and no Campaign A observation is pooled with B. Supplement S3 and S8 show all fourteen candidates; 8/12 is not substituted for the original 8/10 analysis.

Eight usable windows succeeded (Table~\ref{tab:windows}). The exact one-sided binomial tail is
\[
\Pr_{0.5}(X\geq8)=\frac{\binom{10}{8}+\binom{10}{9}+\binom{10}{10}}{2^{10}}=0.0546875.
\]
The two-sided 95\% Clopper--Pearson interval \cite{clopper1934} is $[0.4439,0.9748]$; the one-sided 95\% lower bound is 0.4931. H0 is not rejected and H1-B is \textbf{not established}. Non-rejection also does not establish the null hypothesis. The descriptive fraction 8/10 is not a reliability guarantee of 80\%. A passing safety gate (zero violations in 60 primary P5 records) and untriggered kill criteria do not change the test result or establish the model assumptions.

\begin{table}[htbp]\centering\small
\begin{tabular}{lrrcl}\toprule
Window & Min (s) & Max (s) & R/F & Success\\\midrule
W02/C03 & 595.9 & 673.4 & 5/1 & Yes\\
W03/C04 & 20,789.8 & 20,867.5 & 5/1 & No: deadline\\
W04/C06 & 24,719.5 & 26,885.6 & 6/0 & No: deadline\\
W05/C07 & 633.8 & 714.9 & 6/0 & Yes\\
W06/C08 & 612.6 & 701.1 & 5/1 & Yes\\
W07/C10 & 565.6 & 639.1 & 5/1 & Yes\\
W08/C11 & 645.4 & 730.3 & 6/0 & Yes\\
W09/C12 & 574.2 & 642.9 & 6/0 & Yes\\
W10/C13 & 646.2 & 729.6 & 6/0 & Yes\\
W11/C14 & 636.4 & 710.6 & 6/0 & Yes\\
\bottomrule
\end{tabular}
{}
\caption{All ten usable transitions. Min/max are the six local HOLD-to-service latencies. R/F denotes retained/fallback source counts, including late services. W03 and W04 fail only the deadline condition.}\label{tab:windows}
\end{table}
\begin{figure}[htbp]\centering
\includegraphics[width=.95\linewidth]{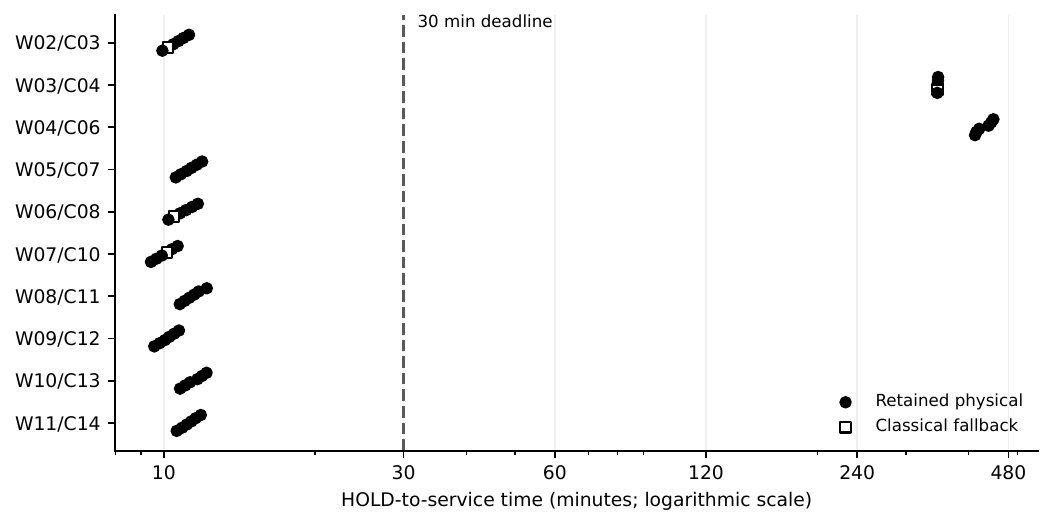}
\caption{Six nested measurements per window, not 60 independent experiments. Points can overlap. The two late windows remain failures; provider waiting is included.}\label{fig:latency}
\end{figure}

The prespecified descriptive leave-one-window-out counts are in Supplement S4. W03/C04 and W04/C06 served correct and valid results after the deadline. Long provider waits, in probe acquisition and in calibration, co-occur with the slow recovery; the observational design does not identify a causal queue effect. At detection, confirmation and pre-opening the queue depths were 2/1/1 for C04 and 1/1/1 for C06. These admitted windows still missed the deadline; admission was not a capacity reservation.

\subsection{Direct checking and the effect of reference renewal}
All 60 held physical probes satisfy the unchanged absolute oracle thresholds and recorded binding requirements. The direct comparator therefore accepts 60/60 without a renewed reference. It also accepts all twelve held probes in C05/C09, assessed separately with information recorded before HOLD; P5 never completes their renewal. These offline decisions are not observations of a deployed service or evidence of its achieved deadline. The comparator checks physical provenance and absolute correctness; P5 additionally demands renewed-reference conformity. Failure to perform that additional contractual step is not evidence of the comparator's inferiority. On this workload, no added correctness or retained-result benefit from renewal is demonstrated.

Table~\ref{tab:direct} gives the joint P2/P5 decisions on the same 60 probes. P2 would retain 57; P5 retains 56 and uses four classical fallbacks. The only changed retention decision is W02/C03 probe 06: its rank decreases from 0.075 to 0.05, so P5 substitutes a fallback for a result P2 would retain. None of P2's three rejections becomes a retained physical result under P5. All four substituted physical results are oracle-correct. They are also the four largest relative errors in this set (approximately 0.0870, 0.0916, 0.1115 and 0.1497), so the rank rule identifies the least-conforming results as designed. The unresolved service question is whether that additional relative-conformity restriction is useful when absolute correctness is already directly checkable. Thus the observed change is narrower than the restoration of eligibility defined by the renewal contract.

\begin{table}[htbp]\centering\small
\begin{tabular}{lrrr}\toprule
P2 replay & P5 retained & P5 fallback & Row total\\\midrule
Serve & 56 & 1 & 57\\
Reject & 0 & 3 & 3\\
\bottomrule
\end{tabular}
{}
\caption{Retrospective decisions on the 60 usable-transition physical probes. The last column is the row total; the direct comparator admits every member of both rows while retaining the same physical result. It makes no renewed-reference claim. These counts do not compare measured deployment latency.}\label{tab:direct}
\end{table}

\begin{figure}[htbp]\centering
\includegraphics[width=\linewidth]{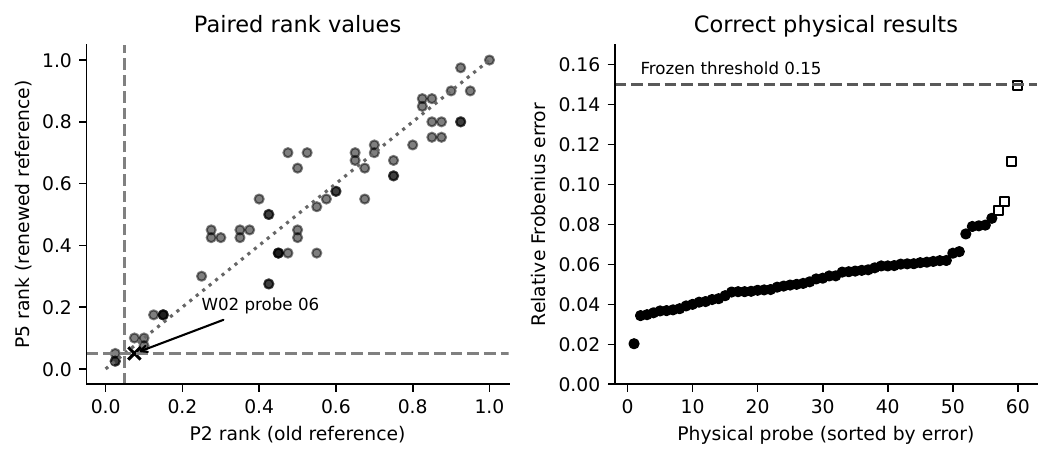}
\caption{Reconstructed ranks and absolute-error margins. The cross identifies the sole changed retention decision. Open squares denote physical results substituted by P5 fallbacks; all are below the frozen relative-error threshold.}\label{fig:comparison}
\end{figure}

The maximum relative error is 0.1496546064, only 0.0003453936 below 0.15; the maximum entrywise absolute error is 0.1555299640, below 0.20. The mean paired rank difference, defined as new minus old, is $-0.004167$, with median zero and range $[-0.175,0.225]$. These summaries do not demonstrate equal reference distributions, absence of physical drift or universal irrelevance of context. Four rank rejections among 60 correct physical results do not validate a 5\% conformal rejection rate: nested probes, sequential acquisition and shared references leave the relevant assumptions unestablished. Full scores, margins and explicitly anchored reference ages are in Online Resource 2.

All 60 primary P5 outputs were oracle-correct and assurance-valid at serving; 48 met the deadline. P4 would withhold all 60 because its support guard rejects the context mismatch. These rule-defined support differences do not demonstrate prevention of a wrong physical answer.

The phase-specific counts and all fourteen descriptive secondary endpoints are retained in Supplement S4. In particular, the canonical aggregate availability values (21/82 for P2 and P4; 70/82 for P5) combine timed P5 primary records with untimed records whose deadline clause defaults to true. They are not a homogeneous latency comparison. All policy records share physical probes; statistical precision cannot be increased by counting the replay policies as additional hardware observations.

The 5/10 retained-only endpoint has nominal Wilson 95\% interval $[0.2366,0.7634]$ \cite{wilson1927}. Four primary fallbacks occurred in W02, W03, W06 and W07; a separate verification fallback occurred in W04. The primary retained fraction is 56/60, not the five-window endpoint: a window can retain all six results and still miss the deadline. The reported probe-level Wilson summaries are descriptive, with within-window dependence unadjusted.

Seven of 582 terminal units failed binding (nominal Wilson 95\% interval $[0.005838,0.024616]$). All seven are C05 calibration units with J3 mismatches. Their concentration and shared conditions leave independence unestablished; they do not estimate an intracluster correlation. C05 is incomplete and excluded from the primary set. The stored zero for \code{windows\_lost\_to\_unbound\_units} counts only accepted windows demoted to non-usable status. It does not mean that no window was affected by binding.

\subsection{Incomplete candidates and limits of observation}
C01 ended before any provider-accepted job, with 28 pre-boundary submission failures. C05 executed 43 provider-DONE jobs: six probes and 37 calibration units. Seven calibrations passed J1, J2 and J4 but failed J3; only 30 calibrations were bound. C05 then terminated at context expiry without a complete reference. C09 also expired before reference completion; the runner requested cancellation of its outstanding job while its recorded status was QUEUED, and subsequently observed CANCELLED (Supplement S3.3). Both C05 and C09 had all six service probes bound and held, but neither completed reference renewal before scientific expiry. Thus held evidence also existed in these two incomplete transitions, beyond the ten usable transitions. This retrospective flow clarification does not reclassify either candidate or change the ten-window contrast. Completion, admissible evidence and usable recovery are different events, so the 582 DONE jobs cannot replace the window denominator.

The C05 chronology was clarified retrospectively (Supplement S6). The seven J3 failures had matching local pre/post snapshots but a different historical projection associated with the provider running timestamp. Eight bound calibration jobs started after the calibration timestamp returned for the unbound group, and another bound job straddled it. Thus neither that timestamp nor the first J3 failure locates a known physical recalibration. The observations establish a disagreement between views and the predicate's response; they do not establish incorrect circuit execution or a physical-state trajectory.

\section{Cost and operational interpretation}
As Rovara et al. discuss, implementing assertions through measurements consumes execution resources \cite{rovara}. Our resource boundary differs: re-evaluation uses zero new service-probe QPU jobs, but recovery is not QPU-free. The ten usable transitions contain 390 calibration jobs, shared across their six probes. This count follows from the fixed reference size and is resource accounting, not an empirical discovery about the renewal rule. The accepted bootstrap adds 39 calibration jobs, and accepted windows include 22 verification jobs. This is 511 jobs for bootstrap plus ten complete transition blocks, before incomplete-window and other acquisition accounting. Calibration is not replicated six times merely because each primary record carries its latency.

\begin{table}[htbp]\centering\small
\begin{tabularx}{\linewidth}{l r Y}\toprule
Quantity & Count/value & Meaning and scope\\\midrule
Primary retained outputs & 56 & Existing results served after reference renewal\\
Primary fallback outputs & 4 & Classical source; not retained-result recovery\\
Recovery calibration jobs & 390 & $39\times10$ usable transitions; shared within window\\
Service-probe reruns after HOLD & 0 & Does not count initial acquisition or calibration\\
Recorded acquisition retries & 1 & C01 calibration; both attempts before provider acceptance\\
Provider-accepted jobs & 583 & 582 DONE and one CANCELLED in closure reconciliation\\
Budget-ledger entries & 611 & Includes 28 C01 pre-boundary SUBMIT\_FAILED entries\\
Provider-reported QPU usage & 1,164 s & Campaign total; not recovery-only or monetary cost\\
Conservative budget usage & 1,833 s & $611\times3$ s accounting estimate\\\bottomrule
\end{tabularx}
\caption{Measured and arithmetically derived resource quantities. No counterfactual cost saving is inferred.}\label{tab:cost}
\end{table}

Across primary probe records, median T0--T3 was 666.026 s, with range 565.562--26,885.649 s. Reference-ready elapsed time T1--T2 ranged from 565.496 to 24,719.410 s across usable windows. These local elapsed times include orchestration and waiting; provider execution durations and provider-reported usage are different measures. The manuscript scripts list the ten distinct calibration latencies rather than treating repeated records as independent calibration runs.

A retrospective timing partition of the two late windows makes the endpoint clock explicit (Supplement S5). C04 and C06 spent 20,789.767 and 24,719.410 s respectively between calibration start and reference readiness. Provider-reported calibration queue intervals summed to 19,606.429 and 23,570.316 s; reported execution intervals summed to 878.703 and 713.662 s. These intervals were non-overlapping and contained within the corresponding local reference interval, but their residual is not identified as exclusively client overhead. Subtracting observed queue intervals would not construct a valid no-queue counterfactual.

C06's complete window lasted 31,445.924 s, and its first HOLD occurred 4,528.160 s after opening. The first probe had already accumulated 4,473.832 s of reported queue time before HOLD. Those seconds are outside that probe's recovery endpoint. After the first HOLD, another 2,166.235 s elapsed before the first calibration submission: the earliest-held probe had already missed the 1,800 s criterion. Sequential acquisition of the remaining probes is therefore a deadline risk distinct from calibration queueing. Reconstructed clocks show that probes 01 and 02 had already exceeded 1,800 s at calibration start. With that observed start time, accelerating only later calibration or re-evaluation cannot recover their deadline. HOLD-to-service must not be advertised as user request-to-response latency, and full-window duration minus maximum recovery latency cannot be assigned entirely to post-service verification.

The direct comparator requires no additional reference jobs once the same bound physical result and exact oracle are available. Renewal adds 39 jobs per complete transition in this implementation. This is a structural algorithmic cost difference, not a measured latency, monetary or energy saving. Classical-only service could instead return the already available exact matrix, but that changes the source and was not a deployed comparison arm. Neither renewal nor physical retention has a demonstrated external necessity here. Any benefit on a different workload or under an independently justified evidence requirement remains to be evaluated.

\section{Discussion and threats to validity}
\subsection{What the study supports}
The defensible decision is conditional: when a retained physical result is bound to the current workload context, a complete reference can be renewed and the result reconsidered without rerunning its service probe. Metadata-consistency binding, operational assurance eligibility and numerical correctness remain separate predicates. B supplies concrete observations of that procedure and of late-but-correct recovery. In relation to workflow recovery \cite{workflow}, the narrower question is whether a renewed reference makes already retained evidence eligible within the declared service deadline; it is not whether retries or checkpoints can keep a computation running. It does not establish that recovery succeeds in more than half of the target windows under the prespecified test. The direct comparator further shows that renewed eligibility supplies no demonstrated additional correctness or retained-result benefit in these data.

The 1,800 s criterion is a scientific endpoint; the implementation can serve after it. A production requirement to discard late answers would need a separate enforcement rule and evaluation. Similarly, zero observed safety violations and a fail-closed endpoint definition do not justify a theorem that every future execution is safe. The code and provider interface impose finite observation and trust boundaries.

\subsection{Sampling and statistical assumptions}
Ten windows are a small sample from one backend and one physical support. Several transitions are temporally close, notably W07--W11 within four hours; common load and calibration processes may make independent, homogeneous Bernoulli trials implausible. The exact test is exact under that model, not robust to arbitrary temporal dependence. Within-window probes share a reference and operating conditions. Neither the primary model nor nominal secondary intervals become validated by integrity checks.

Gate conditioning excludes many operating states, and usability requires complete evidence before expiry. Reporting incomplete candidates is necessary to expose that selection. No unconditional recovery probability, deployment incidence or provider-wide availability can be inferred. The workload is small, has an exact oracle, and lacks the adversarial manipulations or realistic application demands needed to establish broader assurance value. The score itself uses the exact reference; independence of the correctness criterion from the policy's calibration data does not make the oracle scalable.

\subsection{Known implementation and measurement limits}
The frozen analyzer detects KC-B4 (any P5 oracle-incorrect service in an accepted phase) but omits it from the final declaration blockers. The primary safety gate would miss an isolated verification error or an error in an accepted non-usable window. A separately versioned post-campaign correction adds that blocker. Synthetic regressions demonstrate the original and corrected behaviors; both analyzers retain B's non-declaration. This correction was not part of the executed campaign and is not a general production validation (Supplement S7).

Campaign B evaluates the kernel path, not an industrial Fleet Management/routing application or a complete integrated ATHENA-AEGIS system. Provider snapshots, local clocks and histories are imperfect measurements; observed queue association cannot identify causality or justify subtracting delay. No energy, invoice cost, end-user request-to-response comparison or scaled workload arm was collected.

\subsection{Consequences for use}
The evidence supports three concrete reporting and design decisions. First, a service specification should state whether it requires a retained physical result or merely a correct numerical answer; on this workload an exact classical answer is already accessible. Second, its clock must specify the event that starts the deadline and the treatment of acquisition before HOLD, late service and incomplete admissions. Third, a recovery budget must include the shared physical reference, not just the zero-rerun re-evaluation. These decisions are informed by the implementation and observed failures without assuming a queue intervention or a benefit over alternatives. Comparative deployment utility remains unmeasured. Future work should compare a justified renewal requirement against direct checking, test workloads whose correctness evidence is independently useful, and evaluate deadline enforcement and concurrent or anticipatory scheduling. Batch, Session or parallel calibration could change acquisition conditions; these data cannot determine their outcomes. An oracle-free score would require a separate validity argument and evaluation.

\section{Code, data and reproducibility}
Online Resource 2 supplies six count histograms for each of 575 bound physical units, their stored matrices and oracle metrics, all eleven complete 39-score references including bootstrap, incomplete-candidate records, policy decisions and clocks. Its offline command reconstructs matrices, independently regenerates the exact two-qubit reference, recalculates errors and ranks, reconstructs source decisions and endpoints, and computes the binomial tail and intervals from the reconstructed successes. It compares these results with the preserved canonical outputs and generates the figures and tables. The seven unbound calibrators have binding records but no retained count histograms in this export; their missing measurements are not inferred. The artifact also contains the separate KC-B4 correction and synthetic tests. Its editorial wrapper explicitly loads packaged helpers by path, including under Python isolated mode, and regenerates the descriptive context table from the retained projections.

Numerical reproduction starts from locally preserved normalized counts, not independently retrieved provider results. Endpoint reconstruction is conditional on the recorded usable-window set, clocks and the flags for held-result existence, trusted invariants and service-probe reruns. It does not re-execute the complete stratum-support guard or the historical acquisition state machine (Supplement S7.2). J1--J3 comparisons are rechecked from exported timestamps and context digests; recorded integrity/authentication verdicts remain premises. File hashes do not authenticate provider truth, verify private HMAC signatures or independently timestamp a protocol. No provider reconciliation or acquisition is performed by the command. Supplement S7 separates these limits, the historical acquisition environment and the analysis environment.

\paragraph{Manuscript preparation and AI assistance.} OpenAI Codex and Anthropic Claude Code assisted with drafting and language review, literature checking, analysis-code and figure preparation, and internal editorial review. The human authors retain full responsibility for the manuscript.

\section{Conclusion}
The implemented renewal procedure reconsidered retained physical results under a prespecified protocol, with eight successful windows out of ten and five successes using retained results alone. The exact one-sided $p=0.0546875$ does not establish H1-B. Two windows missed the deadline, and renewal required 390 calibration jobs. Direct binding-plus-oracle checking accepts all 60 usable-transition physical results without renewal; renewal changes one retention decision and rescues no P2 rejection as a physical result. These data support an empirical implementation and operational-limit study. Because the exact oracle supplies both scores and the correctness target, this design cannot demonstrate added numerical-correctness utility of renewal over direct checking. The transferable lesson is to justify the reference requirement separately from correctness, expose completion-based selection and source changes, and account for the whole deadline path. No general safety theorem, quantum advantage or deployment superiority follows.

\section*{Statements and Declarations}
\paragraph{Author contributions.} Roberto Fernández-Barrios: Conceptualization, Methodology, Software, Data curation, Formal analysis, Visualization, Investigation, Writing -- original draft, Writing -- review \& editing. Iker Pastor-López: Supervision, Validation, Writing -- review \& editing. Amaia Pikatza-Huerga: Validation, Project administration, Writing -- review \& editing. Pablo García Bringas: Supervision, Resources, Funding acquisition, Writing -- review \& editing.
{}
\paragraph{Funding.} This work is part of grant PID2024-155693NB-C43, ATHENA-AEGIS (Advanced Secure Technologies for Hybrid Quantum-Classical Environments and Applications), funded by MICIU/AEI/10.13039/501100011033 and by ERDF/EU. MICIU and AEI denote the Ministerio de Ciencia, Innovación y Universidades and the Agencia Estatal de Investigación, respectively; ERDF/EU denotes the European Union through the European Regional Development Fund.
\paragraph{Competing Interests.} The authors have no relevant financial or non-financial interests to disclose.
{}
\paragraph{Data and code availability.} Online Resource 2 contains deidentified count histograms, matrices, reference scores, policy records and timestamps, frozen analysis code, a separate corrected analyzer and an offline numerical reproduction command. Ancillary Campaign A and historical timing summaries retain their stated documentary scope. The public reproduction package, including this unchanged numerical artifact, is archived at \url{https://doi.org/10.5281/zenodo.24108573}. Private keys, raw provider identifiers and internal correspondence are excluded. HMAC authentication and provider reconciliation cannot be repeated from this export.

\section*{Supplementary material captions}
\noindent\textbf{Online Resource 1 (ESM\_1.pdf).} Supplementary protocol, campaign populations and outcomes, retrospective timing and metadata clarifications, and the offline reproduction boundary.\\
\textbf{Online Resource 2 (ESM\_2.zip).} Numerical review artifact: deidentified counts and matrices, reference scores, decisions and clocks, offline reconstruction and figures, frozen sources and a separately corrected analyzer, with provenance and integrity manifests.

\section*{Abbreviations}
QPU: quantum processing unit; HMAC: hash-based message authentication code. P2, P4 and P5 denote the policy definitions in Section 4; KC denotes a prespecified kill criterion.
{\interlinepenalty=10000
\bibliography{references}}
\end{document}